\documentclass[usenatbib,fleqn,useAMS]{mn2e}

\usepackage{amsmath}
\usepackage{txfonts}

\renewcommand\pi\upi
\newcommand\dm{\mathrm d}

\title[Tracer Mass Estimators]
{Modified Virial Formulae and the Theory of Mass Estimators}

\author[An \& Evans]
{J.~An$^1$ and N.~W.~Evans$^2$
\\$^1$ National Astronomical Observatories, Chinese Academy of Sciences,
A20 Datun Road, Chaoyang District, Beijing 100012, PR~China;
\\$^2$ Institute of Astronomy, University of Cambridge,
Madingley Road, Cambridge CB3 0HA.}

\date{\sc (to appear in
{\it Monthly Notices of the Royal Astronomical Society})}

\pagerange{\pageref{firstpage}--\pageref{lastpage}}\pubyear{2010}

\begin{document}
\label{firstpage}
\maketitle

\begin{abstract}
  We show how to estimate the enclosed mass from the observed motions
  of an ensemble of test particles.  Traditionally, this problem has
  been attacked through virial or projected mass estimators.  Here, we
  examine and extend these systematically, and show how to construct
  an optimal estimator for any given assumption as to the potential.
  The estimators do not explicitly depend on any properties of the
  density of the test objects, which is desirable as in practice such
  information is dominated by selection effects.  As particular
  examples, we also develop estimators tailored for the problem of
  estimating the mass of the Hernquist or NFW dark matter haloes from
  the projected positions and velocities of stars.
\end{abstract}

\section{Introduction}

Here, we consider the general problem of estimating the enclosed mass
(or equivalently the gravitational potential) from kinematical data on
tracer populations.  In other words, suppose there are $N$ test
particles moving in a gravitational potential $\phi(\bmath r)$
generated by a mass density $\rho(\bmath r)$.  The data available to
us are the instantaneous positions $\bmath r_i$ and velocities $\bmath
v_i$ of the test bodies.  However, it is only rarely that the full
phase space information is available and often only components of
position and velocity along the line-of-sight are measured.  From
these data, we wish to estimate the underlying gravitational potential
or mass by a robust and unbiased statistical method.

This problem has many applications in modern astrophysics -- including
estimating the mass of the Milky Way and M31 from the
kinematics of distant satellites \citep{LT,We99,WEA}, estimating the
mass of the haloes of dwarf galaxies from the stellar velocities
\citep{St08,Wa09,Wo10}, and estimating the mass of galaxy groups and
clusters from their members \citep{He85,Tu06}.  In fact, the
kinematical properties of tracer populations are one of the richest
sources of data on the distribution of dark matter in galaxies and
clusters.  Therefore, it is important to extract as much information
from the data as we possibly can.

Given the significance of the problem, there has been surprisingly
little effort on developing the systematic theory of mass estimators.
Early work \citep{Li60} exploited the virial theorem to obtain
\begin{equation*}
M=\frac{3\pi}{2G}\frac{\langle v_\ell^2\rangle}{\langle R^{-1}\rangle}
\approx\frac{3\pi}{2G}\frac{\sum_i^Nv_{\ell,i}^2}{\sum_i^NR_i^{-1}}
\end{equation*}
for the mass enclosed by $N$ test particles with line-of-sight
velocities $v_{\ell,i}$ and projected positions $R_i$ of each
particle.  The problem with this method was pointed out by
\citet{BT81}, namely that the virial mass estimator is both biased and
inefficient.  These latter authors introduced the alternative
projected mass estimator
\begin{equation*}
M=\frac CG\left\langle v_\ell^2R\right\rangle
\approx\frac C{GN}\sum_i^Nv_{\ell,i}^2R_i
\end{equation*}
where $C$ is a constant determined by the host potential and the
eccentricity of the orbits.

There has also been substantial previous work done on scale-free mass
estimators \citep{Wh81,Ku92,Ev03}.  Recently, \citet*{WEA} formalized
and expanded the ideas from these previous papers, presenting a variety
of mass estimators tailored to scale-free potentials and densities.
These work by taking weighted averages of the combinations of
velocities and positions that remain invariant under similarity
transformations.  None the less, it is also important to devise mass
estimators that are optimized for more realistic and specific
astrophysical potentials.  For example, there are cosmological
arguments that the dark halo density is cusped like $r^{-1}$ at small
radii and falls off like $r^{-3}$ at large radii \citep*{NFW}.  It is
natural to look for mass estimators that build upon this assumption at
the very start.

In this paper, we show how to find such mass estimators tailored for
any given potential.  We also find that our new estimators do not
depend on the number density of the tracers at all.  This is a real
advantage -- e.g., in the case of the Milky Way, the variation
of the number density of the known satellite galaxies and globular
clusters is dominated by the selection effects and the true number
density can only be guessed at.

This paper is arranged as follows.  In Sect.~\ref{sec:th}, we develop
some general theories on mass estimators, showing how to construct one
suitable for given potential.  In Sect.~\ref{sec:ex}, we give a few
specific examples for the cases of astrophysical interest, whilst
Sect.~\ref{sec:proj} sketches the extension to projected data.  The
self-consistent case, when the potential and the density of the
tracers are related through the Poisson equation, is dealt with in
Sect.~\ref{sec:scc}.  Finally, in Sect.~\ref{sec:dis}, we provide a
discussion and conclusions.  Some applications of our estimators to
the widely-used cosmological halo model of \citet{NFW} are found in
\citet*{Ev10}, which should be considered as a companion to
the current paper.

\section{The Theory of Mass Estimators}
\label{sec:th}

\subsection{Jeans' Equation and the Virial Theorem}

Suppose our tracer population has a number density $\nu(r)$ and
a radial velocity dispersion $\sigma_r^2(r)$ and
is moving in a spherical dark halo potential $\phi(r)$,
which by Newton's Theorem satisfies
\begin{equation}
\frac{\dm\phi}{\dm r}=-\frac{GM(r)}{r^2}
\end{equation}
where $M(r)$ is the enclosed halo mass within radius $r$.
These quantities are related to one another through the spherical Jeans
equation that reads
\begin{equation*}
\frac\dm{\dm r}(\nu\sigma_r^2)+\frac{2\beta}r\nu\sigma_r^2
=\nu\frac{\dm\phi}{\dm r}
\end{equation*}
where
\begin{equation*}
\beta=1-\frac{\sigma_\theta^2+\sigma_\phi^2}{2\sigma_r^2}
=1-\frac{\sigma_\theta^2}{\sigma_r^2}
\end{equation*}
is the so-called Binney anisotropy parameter for the spherical system.
The typical application of the Jeans equations involves deriving the
potential and the dark halo mass profile from the observed behaviour
of the tracer density and velocity dispersions (`Jeans modelling').
If the observations are composed of discrete sample datapoints, this
is subject to the uncertainties related to the binning and requires
large number of datapoints to extract any meaningful information. An
alternative when only moderate number of datapoints are available is
to consider the system as whole such as utilizing the virial
theorem. The relation between these two is most obvious in a spherical
system, for which integrating the spherical Jeans equation essentially
results in the scalar virial theorem.

In order to see this, we start by noting that the spherical Jeans
equation reduces to an exact differential form
\begin{equation}\label{eq:jeans}
\frac1Q\frac\dm{\dm r}(Q\nu\sigma_r^2)=-\nu\frac{GM}{r^2}
\end{equation}
by means of the integrating factor $Q=Q(r)$ satisfying
\begin{equation*}
\frac{\dm\ln Q(r)}{\dm r}=\frac{2\beta(r)}r.
\end{equation*}
Next we find that the relation between the local three-dimensional
velocity dispersion and the radial velocity dispersion is given by
\begin{equation}
\sigma^2=\sigma_r^2+\sigma_\theta^2+\sigma_\phi^2
=(3-2\beta)\sigma_r^2=\sigma_r^2\frac{\dm\ln(r^3Q^{-1})}{\dm\ln r}.
\end{equation}
The three-dimensional velocity dispersion of the tracers within
the sphere of the radius of $r_\mathrm{out}$ is thus given by
\begin{equation}
\langle v^2\rangle
=\frac{4\pi}{N_\mathrm{tot}}\int_0^{r_\mathrm{out}}\!\dm r\,r^2\nu\sigma^2
=\frac{4\pi}{N_\mathrm{tot}}\int_0^{r_\mathrm{out}}\!
Q\nu\sigma_r^2\,\frac{\dm(r^3Q^{-1})}{\dm r}\,\dm r
\end{equation}
where
\begin{equation*}
N_\mathrm{tot}=4\pi\int_{r_\mathrm{in}}^{r_\mathrm{out}}\!\dm r\,r^2\nu
\end{equation*}
is the total number of the tracers between an inner $r_\mathrm{in}$
and outer $r_\mathrm{out}$ radius. Integrating by parts, setting
$r_\mathrm{in}=0$ and also using equation (\ref{eq:jeans}) results in
\begin{equation}\label{eq:vir0}\begin{split}
\frac{N_\mathrm{tot}}{4\pi}\langle v^2\rangle
&=\left.\nu\sigma_r^2r^3\right|_0^{r_\mathrm{out}}
-\int_0^{r_\mathrm{out}}\!\frac{r^3}Q\frac{\dm(Q\nu\sigma_r^2)}{\dm r}\,\dm r
\\&=\left.\nu\sigma_r^2r^3\right|_{r=r_\mathrm{out}}
+\int_0^{r_\mathrm{out}}\!\dm r\,GMr\nu,
\end{split}\end{equation}
given that $\sigma_r^2$ is not divergent as $r\rightarrow0$. Here the
last integral actually defines the total potential energy of the
tracers within the same sphere, i.e.,
\begin{equation*}
|W|=4\pi\int_0^{r_\mathrm{out}}\!\dm r\,r^2\nu\frac{GM}r
\,;\qquad
\left\langle\frac{GM}r\right\rangle=\frac{|W|}{N_\mathrm{tot}}.
\end{equation*}
Hence, equation (\ref{eq:vir0}) reduces to
\begin{equation}\label{eq:vir}
\langle v^2\rangle=\left\langle\frac{GM}r\right\rangle
+3\varsigma^2
\end{equation}
where
\begin{equation*}
\varsigma^2
=\frac{\nu(r_\mathrm{out})\,\sigma_r^2(r_\mathrm{out})}
{\bar\nu_\mathrm{out}}
\,;\qquad
\bar\nu_\mathrm{out}=\frac{3N_\mathrm{tot}}{4\pi r_\mathrm{out}^3}.
\end{equation*}
Note that $\bar\nu_\mathrm{out}$ is the mean number density of the
tracers in the sphere of the radius of $r_\mathrm{out}$.

Equation (\ref{eq:vir}) is in fact equivalent to the statement of the
scalar virial theorem for a pressure-supported spherical system as
$\frac12\langle v^2\rangle$ is basically the kinetic energy per tracer
particle associated with the random motion. The presence of the
boundary term (that is, the surface term, $3\varsigma^2$)
is due to the hard cut-off
at $r=r_\mathrm{out}$, which can correspond to the situation when the
tracers are confined within the spherical radius of $r_\mathrm{out}$
through the external pressure. However, it is usual to drop the
boundary term if the system as a whole is considered.

\subsection{Tracer Mass Estimators}

The virial theorem (eq.~\ref{eq:vir}) is traditionally used to
estimate the total mass of the system. The integral mean value theorem
indicates that there exists a kind of `mean radius' $\bar r$ within
the interval bounded by the outer cut-off $r_\mathrm{out}$ (i.e.,
$0<\bar r\le r_\mathrm{out}$) such that $\langle GM/r\rangle=GM(\bar
r)/\bar r$. Therefore, if one ignores the boundary term, one can
relate the mass within the `mean radius' to the total velocity
dispersion of the tracers, $M(\bar r)=\bar r\langle v^2\rangle/G$,
which may also be suspected from a simple dimensional analysis. If the
distribution of the gravitating mass is known or assumed, the
definition of $\bar r$ can be made precise and furthermore $M(\bar r)$
can be scaled to provide the estimate of $M(r_\mathrm{out})$. That is
to say, we can relate the total mass of the system to the certain
integrals of kinetic properties of the tracers.

However, this approach suffers from drawbacks related to the fact that
the mass estimate depends on two separate averages \citep[see
e.g.,][]{BT81}. This difficulty is partially overcome by the use of
`mass estimator', that is, an average of particular combinations of
kinetic properties of the tracers that directly relates to the total
mass rather than to the potential energy, as does the virial
estimator. However, derivation of the proper form of the mass
estimator requires some analysis of the dynamics of the system. Here
we still consider the simplest case of the spherical system traced by
a non-rotating relaxed population in equilibrium.

First we note that the virial theorem (with the boundary term dropped)
indicates that $\langle v^2\rangle/\langle v_\mathrm c^2\rangle=1$ where
$v_\mathrm c=(GM/r)^{1/2}$ is the circular speed of the potential.
From this, one may naively expect that
$\langle v^2/v_\mathrm c^2\rangle\approx 1$, but the distributed mass
and tracers only make this approximately so. However, with a proper
weighting $f(r)$, we can actually show that there exists a relation
$\langle fv_r^2/v_\mathrm c^2\rangle=1$, which will be subsequently
used to derive a proper mass estimator.

Let us assume for the moment that the spherical dark halo profile
$M(r)$ is known. Then we can show that the proper weighting function
is given by
\begin{equation}\label{eq:coef}\begin{split}
f(r)&=4-2\beta(r)-\frac{\dm\ln M(r)}{\dm\ln r}
\\&=\frac{\dm\ln(r^4Q^{-1}M^{-1})}{\dm\ln r}
=\frac{QM}{r^3}\frac\dm{\dm r}\biggl(\frac{r^4}{QM}\biggr).
\end{split}\end{equation}
Then the `weighted' average of the tracer radial velocities $v^2_r$ in
a spherical system given by
\begin{equation}\begin{split}
\left\langle\frac{fv_r^2}{v_\mathrm c^2}\right\rangle
&=\frac{4\pi}{N_\mathrm{tot}}
\int_{r_\mathrm{in}}^{r_\mathrm{out}}\frac{f\nu\sigma_r^2r^3}{GM}\,\dm r
\\&=\frac{4\pi}{GN_\mathrm{tot}}
\int_{r_\mathrm{in}}^{r_\mathrm{out}}
Q\nu\sigma_r^2\frac\dm{\dm r}\biggl(\frac{r^4}{QM}\biggr)\,\dm r.
\end{split}\end{equation}
Integrating by part leads to
\begin{equation}\begin{split}
\frac{GN_\mathrm{tot}}{4\pi}
\left\langle\frac{fv_r^2}{v_\mathrm c^2}\right\rangle
&=\left.\frac{\nu\sigma_r^2r^4}M\right|_{r_\mathrm{in}}^{r_\mathrm{out}}
-\int_{r_\mathrm{in}}^{r_\mathrm{out}}
\frac{r^4}{QM}\frac\dm{\dm r}(Q\nu\sigma_r^2)\,\dm r
\\&=\left.\frac{\nu\sigma_r^2r^4}M\right|_{r_\mathrm{in}}^{r_\mathrm{out}}
+\frac{GN_\mathrm{tot}}{4\pi},
\end{split}\end{equation}
where we have also used equation (\ref{eq:jeans}).

For any physical system, we apply the result of \citet{AE09} to find
that $\lim_{r\rightarrow0}r^4\nu\sigma_r^2/M=0$.\footnote{Naively, we
have $\lim_{r\rightarrow0}r^3\nu=0$ for any physical $\nu$ since
otherwise there would be an infinite mass concentration of the tracers
at the centre whereas $\lim_{r\rightarrow0}r\sigma_r^2/M$ is
typically finite \citep{AE09}. More careful examination of
\citet{AE09} indicates that even for the exceptional case that
$\lim_{r\rightarrow0}r\sigma_r^2/M$ diverges, the boundary term
still vanishes as $\lim_{r\rightarrow0}r^3\nu=0$ is always
dominant. That is to say, we infer from \citet{AE09} that, given $M(r)$
behaving as $\sim r^{1-\alpha}$ as $r\rightarrow0$, $\alpha>2\beta-3$
is the sufficient condition for this. However, if $\nu\sim r^{-\gamma}$
as $r\rightarrow0$, we have $2\beta\le\gamma\le\alpha+2$ -- the first
inequality is due to \citet{AE06} and the second to the fact that the
tracers cannot be cusped steeper than the dark halo -- and so it is
met.} Hence, setting $r_\mathrm{in}=0$ leads to the vanishing inner
boundary term, and thus
\begin{equation}
\left\langle\frac{fv_r^2}{v_\mathrm c^2}\right\rangle
=1+3\,\frac{\varsigma^2}{v_\mathrm c^2(r_\mathrm{out})}.
\end{equation}
This relation can be rearranged to yield the total dark halo mass
$M_\mathrm{out}=M(r_\mathrm{out})$ within radius $r_\mathrm{out}$, 
\begin{equation}\label{eq:main}
GM_\mathrm{out}
=\left\langle\frac{frv_r^2}{\tilde\mu}\right\rangle
-3r_\mathrm{out}\varsigma^2,
\end{equation}
where
\begin{equation}\label{eq:mprof}
\tilde\mu(r)=\frac{M(r)}{M_\mathrm{out}}
\qquad\mbox{($0\le r\le r_\mathrm{out}$)}
\end{equation}
is the normalized dark halo mass profile function. Note that
$\dm\ln\tilde\mu/\dm\ln r=\dm\ln M/\dm\ln r$,
and thus $f(r)$ can be evaluated
if $\tilde\mu(r)$ is known without any reference to $M_\mathrm{out}$.
Consequently, if we are willing to assume the functional form of the
halo mass profile $\tilde\mu(r)$, which is normalized to be
$\tilde\mu(r_\mathrm{out})=1$, in a spherical region of interest,
$r<r_\mathrm{out}$, then the total halo mass $M_\mathrm{out}$ within
the same spherical region can be estimated through a particular
average of kinematic properties of the tracers, which is in practice
inferred from the corresponding discrete sample mean, i.e.,
\begin{equation}
M_\mathrm{out}\approx
\frac1{GN}\sum_i^N\frac{r_if(r_i)}{\tilde\mu(r_i)}v_{r,i}^2,
\end{equation}
possibly further adjusted by the boundary term
($3r_\mathrm{out}G^{-1}\varsigma^2$) if necessary.

Here, the presence of the boundary term is again related to the
external pressure support of the tracer population. If the tracer
population is a true isolated system of a finite spherical extent of
$r_\mathrm{out}$ in equilibrium with the dark halo potential, it
follows that $\sigma_r^2(r_\mathrm{out})=\nu(r_\mathrm{out})=0$ and
the boundary term naturally vanishes. A similar argument extends to
the system of an infinite-extent tracer population with a
finite-total-mass dark halo, for which $r_\mathrm{out}=\infty$ also
leads to the dropped outer boundary term (then
$M_\mathrm{out}=M_\mathrm{tot}$ is now the `true' total halo mass).
However, if the tracer population is pressure-confined and/or the
distribution of the observed tracers are truncated at a finite outer
radius $r_\mathrm{out}$, the outer boundary term must remain. In this
case, if enough datapoints are available, the boundary term may be
directly calculated from the observed tracer distribution.

\section{Examples}
\label{sec:ex}

We now develop formulae specific to some simple and widely-used
halo models.

\subsection{The Scale-Free Potential}

The simplest mass model that we consider is
($0\le r\le r_\mathrm{out}$)
\begin{equation}\label{eq:musf}
\tilde\mu(r)=\biggl(\frac r{r_\mathrm{out}}\biggr)^{1-\alpha}.
\end{equation}
Here, $\alpha\le1$ is the power-index for the scale-free potential,
or equivalently the rotation curve is given by
\begin{equation*}
v_\mathrm c^2(r)=
\frac{r_\mathrm{out}^\alpha}{r^\alpha}\,v_\mathrm c^2(r_\mathrm{out}).
\end{equation*}
The central point-mass case is included with $\alpha=1$. If we
require that the halo density does not increase outwards, then the
index is restricted to be $\alpha\ge-2$, with $\alpha=-2$
corresponding to a homogeneous sphere.

For these cases, we have
$f(r)=3-2\beta+\alpha$. If $\beta$ is further assumed to be constant,
then $f$ is also constant, and therefore equation
(\ref{eq:main}) reduces to
\begin{equation}\label{eq:ssp}
\frac{GM_\mathrm{out}}{r_\mathrm{out}}
=(\alpha+3-2\beta)
\frac{\langle v_r^2r^\alpha\rangle}{r_\mathrm{out}^\alpha}
-3\varsigma^2.
\end{equation}
These are similar to the estimators used by \citet{WEA} but the
multiplicative coefficient $\alpha+\gamma-2\beta$ in their equation
(15) is replaced by $\gamma\Rightarrow3$. This is because \citet{WEA}
further assumed a power-law behaviour for the tracer density profile,
i.e., $\nu\propto r^{-\gamma}$, which allowed them to solve for
$N_\mathrm{tot}$ and $\sigma_r^2(r)$. Instead of dropping the boundary
term by sending $r_\mathrm{out}\rightarrow\infty$, which is an
improper thing to do in a scale-free system, they equated $4\pi
r^3\nu/N(r)\approx3-\gamma$ and
$\sigma_r^2\simeq(\gamma-2\beta+\alpha)^{-1}GM/r$ based on the
power-law solutions. If we substitute these for the boundary term in
equation (\ref{eq:ssp}), we recover their equation (15).

Among the scale-free potential cases, of particular interest are the
Kepler ($\alpha=1$) and the logarithmic ($\alpha=0$) potential. For
the Kepler potential generated by a central point mass, it is usual
to set $r_\mathrm{out}=\infty$ and $\varsigma^2=0$. Given that $\beta$
is also constant, this leads to
\begin{equation}\label{eq:kep}
\langle v_r^2r\rangle=\frac{GM_\bullet}{2(2-\beta)},
\end{equation}
where $M_\bullet$ is now the mass of the central point.
On the other hand, for the logarithmic potential generated by a
(truncated) singular isothermal sphere, equation (\ref{eq:ssp})
reduces to
\begin{equation*}
v_\mathrm c^2=\langle v^2\rangle
-3\varsigma^2,
\end{equation*}
which is actually the same as equation (\ref{eq:vir}) with
$\langle GM/r\rangle=v_\mathrm c^2$ being constant.

\subsection{The Double-Power Law Halo}

A widely-used family to fit the simulated dark halo density profiles
is in the form of $\rho(r)\propto r^{-a}(r_0^p+r^p)^{-(b-a)/p}$. If
$b>3$, the total dark halo mass $M_\mathrm{tot}$
is finite and $\tilde\mu(r)$ reduces
to the regularized beta function with $r_\mathrm{out}=\infty$.
However, for many specific cases the results are much simpler. For
example, if $[\rho(r)]^{-1}\propto r^\gamma(r_0^p+r^p)^{(3-\gamma)/p+1}$
with $0\le\gamma<3$ and $p>0$, then
\begin{equation}\label{eq:dpm}
\frac{M(r)}{M_\mathrm{tot}}
=\left(1+\frac{r_0^p}{r^p}\right)^{-(3-\gamma)/p};\
f=4-2\beta-\frac{3-\gamma}{(r/r_0)^p+1}.
\end{equation}
That is, the functions $\tilde\mu(r)$ and $f(r)$ are in easily
tractable analytic form provided that $\beta(r)$ is as such. This
particular example includes the well-known families of the
$\gamma$-sphere \citep{De93,Tr94} with $p=1$ and that of \citet{Ve79}
and \citet{EA05} with $\gamma=2-p$.

In practice, the simple analytic form of $\tilde\mu(r)$ and $f(r)$
indicates that the sample mean of $\tilde\mu^{-1}frv_r^2$ is
straightforward to calculate for a fixed set of parameters. Moreover,
provided that the tracer population extends sufficiently far out (that
is, $r_\mathrm{out}\gg r_0$) and so
$M_\mathrm{out}\equiv M(r_\mathrm{out})\approx M_\mathrm{tot}$,
we also argue that it is in general safe to drop the
boundary term (formally $r_\mathrm{out}\rightarrow\infty$, and
$\tilde\mu=M/M_\mathrm{tot}$).

For instance, for the \citet{He90} halo profile, that is,
$p=\gamma=1$ in equation (\ref{eq:dpm}), we find that
\begin{equation}\begin{split}
GM_\mathrm{tot}
&=\left\langle
\biggl(4-2\beta-\frac2{1+r/r_0}\biggr)\,
\biggl(1+\frac{r_0}r\biggr)^2rv_r^2\right\rangle
\\&\approx\frac{2r_0^2}N\sum_i^N
\left[1+\frac{2r_i}{r_0}-\biggl(1+\frac{r_i}{r_0}\biggr)\,\beta_i\right]
\left(1+\frac{r_i}{r_0}\right)
\frac{v_{r,i}^2}{r_i}
\end{split}\end{equation}
where $\beta_i=\beta(r_i)$. If $r_i\ll r_0$, the observable
combination contributes like $\sim2(1-\beta_i)r_0^2v_{r,i}^2r_i^{-1}$
whereas for $r_i\gg r_0$ the contribution is like
$\sim2(2-\beta_i)v_{r,i}^2r_i$, which is consistent with the local
behaviour of the potential at the location of the tracer particle. we also
note that the mass estimate is dependent upon the choice of the scale
length $r_0$, which is expected.

\subsection{The NFW Halo}

The formal NFW profile \citep{NFW}, i.e.,
$\rho^{-1}\propto r(r_0+r)^2$, on the other hand possesses
infinite total mass, and the proper application of our scheme calls
for the truncation of the profile either at the radius of the
outermost tracer point $r_\mathrm{out}$,
or at the virial radius $r_\mathrm v$. Let us suppose that
the tracers are well-populated so that the mass up to the virial
radius $M_\mathrm v\equiv M(r_\mathrm v)$, i.e., the virial mass
can be effectively estimated. Since $M(r)\propto m(r/r_0)$ where
\begin{equation*}
m(x)=\ln(1+x)-\frac x{1+x}
\end{equation*}
for the NFW profile, if we let $M_\mathrm{out}=M_\mathrm v$ and
$r_\mathrm{out}=r_\mathrm v$, then
\begin{equation}
\tilde\mu(r)=\frac{m(\tilde r)}{m(c)}
\,;\qquad
f(r)=4-2\beta-\frac1{m(\tilde r)}
\left(\frac{\tilde r}{1+\tilde r}\right)^2
\end{equation}
where $c=r_\mathrm v/r_0$ is the concentration parameter and
$\tilde r=r/r_0=cr/r_\mathrm v$.
For a fixed $r_0$, equation (\ref{eq:main}) is scaled to
\begin{equation}\label{eq:nfw}
\frac{GM_\mathrm{out}}{r_0m(c)}
=\left\langle v_r^2\tilde h(\tilde r)\right\rangle
-\frac{3c}{m(c)}\,\varsigma^2
\end{equation}
where $\tilde h(\tilde r)\equiv\tilde rf/m(\tilde r)$ whose
$r$-dependence is only via the scaled radius $\tilde r$.
That is to say,
the $c$-dependence of the result is essentially through the overall
scale. Here, the boundary term is expected to be small\footnote{For a
virialized system, it is expected that $\sigma_r^2(r_\mathrm v)\approx 0$.
In addition, if $r_\mathrm v=r_{200}$, then
$4\pi r_\mathrm v^3\nu(r_\mathrm v)/N_\mathrm{tot}\le
4\pi r_\mathrm v^3\rho(r_\mathrm v)/M_\mathrm{out}\la3/200$ for the
tracers that are more centrally concentrated than the dark halo. With
typical concentration parameter values, then
$3\varsigma^2c/m(c)\la0.1\sigma_r^2(r_\mathrm v)$.}
and so it is all right to ignore it at the given level
of the precision.

\section{Mass Estimators for Projected Data}
\label{sec:proj}

\subsection{Line-of-Sight Velocity Data}

In many situations, the radial velocities ($v_r$) -- with respect to
the centre of the halo -- of the tracers are not direct observables,
but the line-of-sight velocities ($v_\ell$) are. Fortunately, the
adjustment of the estimator relating to the alternative velocity
projections is straightforward. We use the relationship between the
line-of-sight velocity dispersion ($\sigma_\ell$) and the radial one
($\sigma_r$);
\begin{equation*}
\sigma_\ell^2=(1-\beta\sin^2\!\varphi)\,\sigma_r^2
\end{equation*}
where $\varphi$ is the angle between the line of sight towards the
tracer and the radial position vector of the same tracer from the
centre of the halo. If the halo is sufficiently far away from us, each
line of sight towards the individual tracer star or satellite galaxy
runs approximately parallel to the line of sight towards the halo
centre. Then, the angle $\varphi$ is equivalent to the spherical polar
angular coordinate $\theta$ centred on the halo centre. In a
spherical system, $\sigma_r^2$ is only dependent on $r$ and thus by
averaging $\sigma_\ell^2$ over the angle $\theta\approx\varphi$ at a
fixed $r$, we find that
\begin{equation*}
\int_0^{\pi/2}\!\dm\varphi\sin\varphi\,(1-\beta\sin^2\!\varphi)\,\sigma_r^2
=\biggl[1-\frac23\,\beta(r)\biggr]\,\sigma_r^2.
\end{equation*}
Hence the weighted averages of $v_\ell^2$ and $v_r^2$ are related
to each other such that
\begin{equation}\label{eq:vlos}
\left\langle\frac{3v_\ell^2h(r)}{3-2\beta}\right\rangle
=\left\langle v_r^2h(r)\right\rangle
\end{equation}
for any radial weighting function $h(r)$. This is valid for any
$\beta(r)$ provided that the spherical symmetry assumption holds. For
the constant $\beta$ cases, the factor $(1-\frac23\beta)$ is a simple
multiplicative constant that can be applied after the averaging.

Finally, the proper form of the mass estimator involving the
line-of-sight velocities is obtained after substituting equation
(\ref{eq:vlos}) into equation (\ref{eq:main}),
\begin{equation}\label{eq:vlos2}
\frac{GM_\mathrm{out}}3
=\left\langle\frac{3-2\beta+\hat\alpha}{3-2\beta}
\frac{v_\ell^2r}{\tilde\mu}\right\rangle
-r_\mathrm{out}\varsigma^2
\end{equation}
where $\hat\alpha(r)\equiv1-(\dm\ln\tilde\mu/\dm\ln r)$, which would be
a constant $\hat\alpha=\alpha$ if $\tilde\mu$ is given by equation
(\ref{eq:musf}).

\subsection{Projected Separation Data}
\label{sec:prosep}

In more typical cases for an external dark halo, we may only have the
line-of-sight velocities and the projected distance ($R=r\sin\theta$) to
the halo centre known to within a reasonable precision. Here, we would
like to find the proper weighting function $w(R)$ of $R$ such that
$\langle v_\ell^2w(R)\rangle=\langle\tilde\mu^{-1}frv_r^2\rangle$,
which would replace equation (\ref{eq:main}) with
\begin{equation*}
GM_\mathrm{out}
=\left\langle v_\ell^2w(R)\right\rangle-3r_\mathrm{out}\varsigma^2,
\end{equation*}
and lead to the projected mass estimator
\begin{equation*}
M_\mathrm{out}\approx\frac1{GN}\sum_i^Nw(R_i)\,v_{\ell,i}^2.
\end{equation*}
In practice, the average here would be over the cylindrical region
with $R\le R_\mathrm{out}$ whereas the one in equation (\ref{eq:main})
is over the spherical region of $r\le r_\mathrm{out}$. The distinction
is moot if the average is in fact over the whole space, i.e.,
$r_\mathrm{out}=R_\mathrm{out}=\infty$. With a finite cut-off radius
$R_\mathrm{out}$ in the tracer population, we proceed by assuming the
true three-dimensional distribution of tracers is spherically
symmetric and also cuts off at $r_\mathrm{out}=R_\mathrm{out}$. That
is to say, the observed sample mean of $\langle v_\ell^2 w(R)\rangle$
is considered to contain no contribution from tracers with
$r>r_\mathrm{out}$ and therefore to be a practical estimator for the
average in the sphere of radius $r_\mathrm{out}$.

Provided that the system is spherical and both averages are over the
sphere, the condition that $\langle v_\ell^2 w(R)\rangle=\langle
v_r^2h(r)\rangle$ results in a integral equation for $w(R)$ at a fixed
$r$;
\begin{equation}\label{eq:gie}\begin{split}
rh(r)
&=r\int_0^{\pi/2}\!\dm\theta\sin\theta\,
(1-\beta\sin^2\!\theta)\,w(r\sin\theta)
\\&=\int_0^r\left[1-\beta(r)\frac{R^2}{r^2}\right]
\frac{w(R)\,R\,\dm R}{(r^2-R^2)^{1/2}}.
\end{split}\end{equation}
For $h=rf/\tilde\mu$, this is in principle invertible for $w(R)$ if
$\lim_{r\rightarrow0}(\dm\ln\tilde\mu/\dm\ln r)<3$ and $\beta(r)$ is
finite. We refer the reader to Appendix \ref{sec:app} for details.

In particular, if $\beta$ is a finite constant and $\tilde\mu$ is
given by the scale-free form in equation (\ref{eq:musf}) with
$-2<\alpha\le1$, we find that
\begin{equation}
w(R)=
\frac{R^\alpha}{\hat I_{\alpha,\beta}r_\mathrm{out}^{\alpha-1}}
\end{equation}
and
\begin{equation*}
\hat I_{\alpha,\beta}
=\frac{\pi^{1/2}\Gamma\bigl(\tfrac\alpha2+1\bigr)}
{4\Gamma\bigl(\tfrac{\alpha+5}2\bigr)}
\frac{\alpha+3-(\alpha+2)\beta}{\alpha+3-2\beta}
\end{equation*}
where $\Gamma(x)$ is the gamma function (the generalized factorial).
That is to say, the corresponding mass estimator is in the form of
\citep[c.f.,][eqs.~26 \& 27]{WEA}
\begin{equation}\label{eq:wea}
\frac{GM_\mathrm{out}}{r_\mathrm{out}}=
\frac{\langle v_\ell^2R^\alpha\rangle}{\hat I_{\alpha,\beta}r_\mathrm{out}^\alpha}
-3\varsigma^2.
\end{equation}
\citet{BT81} considered the projected mass estimator for the central
point-mass case. Their results are consistent with
equation (\ref{eq:wea}) with $\alpha=1$ once we drop the boundary term
by setting $r_\mathrm{out}=\infty$, that is,
\begin{equation}\label{eq:BT}
GM_\bullet=
\frac{32}\pi\frac{2-\beta}{4-3\beta}
\langle v_\ell^2R\rangle.
\end{equation}
Equation (\ref{eq:wea}) for $\alpha=0$ on the other hand results in
\begin{equation}\label{eq:Wo10}
\langle v_\ell^2\rangle-\frac\nu{\bar\nu_\mathrm{out}}\,\sigma_r^2
=\frac{v_\mathrm c^2}3.
\end{equation}
That is to say, if the rotation curve of the spherical halo is flat,
the line-of-sight velocity dispersion is related to the circular
velocity (and thus the mass) of the spherical halo such that $3\langle
v_\ell^2\rangle\approx v_\mathrm c^2$ \citep[c.f.,][]{LB81,Ev97,Wo10}
independent of the behaviour\footnote{Eq.~(\ref{eq:Wo10}) is in fact
valid even if $\beta$ varies radially. This is because
$\langle v^2\rangle=3\langle v_\ell^2\rangle$ for any spherical system
independent of the behaviour of $\beta$, and therefore
eq.~(\ref{eq:vir}) is equivalent to eq.~(\ref{eq:Wo10}) for the halo
with a flat rotation curve.} of $\beta$, to an extent that one can
ignore the boundary term.

For more complicated mass profiles, an analytic result is in general
difficult to obtain, except for some special cases. However, the
special cases do include some interesting examples, one of which is
the Hernquist halo profile traced by the populations with constant
$\beta$, for which
\begin{equation}
\frac{GM_\mathrm{tot}}{r_0}=\left\langle\biggl(
\frac{32}\pi\frac{2-\beta}{4-3\beta}\frac R{r_0}+6
+\frac8\pi\frac{1-\beta}{2-\beta}\frac{r_0}R
\biggr)\,v_\ell^2\right\rangle
\end{equation}
where $r_0$ is the scale length of the Hernquist halo. The proof is
given in Appendix \ref{app:hern}, together with further examples of
mass profiles that result in a rational projected mass estimator.

Strictly speaking, the Hernquist halo cannot be traced by a population
with $\beta>\frac12$ at the centre in equilibrium since such a
population must be cusped at the centre steeper than the dark halo
cusp, which behaves as $\sim r^{-1}$ \citep{AE06}. However, the
constant anisotropy needs not extend to the centre and the mass
estimate with varying anisotropy in the interval $(-\infty,1]$ may be
understood to be the range of the halo mass consistent with the
observed line-of-sight velocity data set.

This projected mass estimator for the Hernquist profile is again
notably consistent with the scale-free case of equation (\ref{eq:wea})
at either extreme, $R\gg r_0$ (a finite total mass; $\alpha=1$) or
$R\ll r_0$ ($r^{-1}$ cusp; $\alpha=-1$). This indicates that if the
tracers are restricted locally in the region where the halo potential
can be approximated as a power law, equation (\ref{eq:wea}) is a
reasonable proxy for the mass estimator given proper boundary terms.

For general mass profiles, the weighting function can be derived
numerically provided that
$\lim_{r\rightarrow0}(\dm\ln\tilde\mu/\dm\ln r)<3$
(i.e., a cusped halo density profile). While the solution for the
general case involves a double integral at the least, the function
$w(R)$ for a few particular cases of constant $\beta$ with an analytic
mass profile can be obtained through a simple quadrature (see
Appendix \ref{sec:app}). For example, the normalized weighting
functions $\tilde W(\tilde R)$ for the NFW profile -- which replaces
$\tilde h(\tilde r)$ in eq.~\ref{eq:nfw} together with the change to
the line-of-sight velocity and $\tilde R=R/r_0$ -- for some constant
$\beta$ are provided in figure~1 of \citet{Ev10}. For the
particular case of the NFW profile, it is also possible to derive
the power-series expansion of $\tilde W(\tilde R)$ at $\tilde R=0$
analytically, and also its
asymptotic behaviour towards $\tilde R\rightarrow\infty$.
In particular, we find at $\tilde R=0$ that
$\tilde W(\tilde R)\simeq
\frac{16}\pi C_\beta\tilde R^{-1}+8+\mathcal O(\tilde R)$ where
$C_\beta=(1-\beta)/(2-\beta)$ and towards $\tilde
R\rightarrow\infty$, that $\tilde W(\tilde R)\sim
\tilde R/[\ln(1+\tilde R)-1]\sim\tilde R/(\ln\tilde R)$.

Finally, the circular orbit model ($\beta=-\infty$) has some special
points of interest, which are discussed in Appendix \ref{app:cir}. 

\subsubsection{Cored Halo Profiles}

Equation (\ref{eq:wea}) is invalid for $\alpha=-2$ (i.e., an
homogeneous sphere of radius $r_\mathrm{out}$) because, provided
that $v_\ell^2\ne0$ for $R=0$ (i.e., $\beta\ne-\infty$), the average
$\langle v_\ell^2/R^2\rangle$ that extends to $R=0$ diverges. However,
since $\hat I_{\alpha=-2,\beta>-\infty}$ also diverges, it actually
leads to an indeterminate form. In fact, the formal solution for
equation (\ref{eq:gie}) with $\tilde\mu=(r/r_\mathrm{out})^3$ exists
in that $w(R)=2(1-2\beta)r_\mathrm{out}^3\delta(R^2)$ where
$\delta(x)$ is the Dirac delta -- that is to say, letting
$R^\alpha/\Gamma(\frac\alpha2+1)\rightarrow\delta(R^2)$ as
$\alpha\rightarrow2$. In practice, the average $\langle
v_\ell^2\delta(R^2)\rangle$ is directly related to the
tracer-number-weighted line-of-sight velocity dispersion along the
central line of sight, $\sigma_\mathrm{L,0}$, or,
\begin{equation}
\left\langle v_\ell^2\delta(R^2)\right\rangle
=\frac{2\pi}{N_\mathrm{tot}}\int\!\dm r\,\nu\sigma_r^2
=\frac\pi{N_\mathrm{tot}}\Sigma_0\sigma_\mathrm{L,0}^2
\end{equation}
where $\Sigma_0$ is the tracer column density along the central line
of sight. Therefore, for $\alpha=-2$ (and $\beta\ne-\infty$), equation
(\ref{eq:wea}) is replaced by
\begin{equation}
\frac{GM_\mathrm{out}}{r_\mathrm{out}}
=2(1-2\beta)\frac{\Sigma_0\sigma_\mathrm{L,0}^2}{\bar\Sigma_\mathrm{out}}
-3\varsigma^2
\end{equation}
where $\bar\Sigma_\mathrm{out}=N_\mathrm{tot}/(\pi r_\mathrm{out}^2)$
is the mean column density of the total tracer population.

It can also be shown that the properly derived weighting function
$w(R)$ for any cored halo mass profile with
$\lim_{r\rightarrow0}(\dm\ln\tilde\mu/\dm\ln r)=3$ contains the Dirac
delta. Specifically, if $\lim_{r\rightarrow0}r^3/\tilde\mu=L$ is a
finite nonzero, then replacing $h$ with $\hat h=h-2(1-2\beta)L/r^2$
allows equation (\ref{eq:gie}) to be inverted. If $\hat w(R)$ is the
solution for this inversion, the final weighting function that
reproduces $\langle v_r^2h(r)\rangle=\langle v_\ell^2 w(R)\rangle$ is
found to be $w(R)=2(1-2\beta)L\delta(R^2)+\hat w(R)$. That is to say,
we find that
\begin{equation}
\left\langle v_\ell^2\hat w(R)\right\rangle
=\frac{GM_\mathrm{out}}{r_\mathrm{out}}
-\frac{2(1-2\beta)L}{r_\mathrm{out}^3}
\frac{\Sigma_0\sigma_\mathrm{L,0}^2}{\bar\Sigma_\mathrm{out}}
+3\varsigma^2
\end{equation}
given that
\begin{equation*}
\frac1r\left[\frac{r^3f}{\tilde\mu}-2(1-2\beta)L\right]
=\int_0^r\left(1-\beta\frac{R^2}{r^2}\right)
\frac{\hat w(R)\,R\,\dm R}{(r^2-R^2)^{1/2}}
\end{equation*}
and $L=\lim_{r\rightarrow0}r^3/\tilde\mu$ is a finite constant (which
further implies that $\lim_{r\rightarrow0}r^3f/\tilde\mu=2(1-2\beta)L$
and thus the above integral equation is invertible).

\section{The Self-Consistent Case}
\label{sec:scc}

If the tracer density $\nu(r)$ follows the same functional form as the
dark halo density $\rho(r)$ (i.e., the mass-to-light is constant),
then $\tilde\mu(r)$ is specified by the integral of $\nu(r)$ over the
volume, and so the problem is completely determined. However, for this
case the problem can be approached through a different simpler route,
directly utilizing the fact that the potential and the tracer density
are related through the Poisson equation. We find that the
self-consistent case result in a formally identical mass estimator as
the point-mass case except for an exact factor of two difference in
the associated constant.

If $\nu/\rho$ is constant, using $\dm M/\dm r=4\pi r^2\rho$,
equation (\ref{eq:jeans}) reduces to
\begin{equation*}
\frac{\dm M^2}{\dm r}
=-\frac{8\pi}G\frac{r^4}Q\frac\dm{\dm r}(Q\rho\sigma_r^2).
\end{equation*}
Hence, integrating this on $r$ over $[0,\infty)$, we find that
\begin{equation*}
\frac{GM_\mathrm{tot}^2}{8\pi}
=-\int_0^\infty\frac{r^4}Q\frac\dm{\dm r}(Q\rho\sigma_r^2)\,dr
=\int_0^\infty Q\rho\sigma_r^2
\frac\dm{\dm r}\biggl(\frac{r^4}Q\biggr)\,\dm r
\end{equation*}
where $M_\mathrm{tot}=M(\infty)$ is the total mass. That is to say,
\begin{equation}\label{eq:scc}
M_\mathrm{tot}=\frac2G\left\langle(4-2\beta)\,v_r^2r\right\rangle.
\end{equation}
The result is valid for an arbitrary functional form for the anisotropy
parameter $\beta=\beta(r)$, but if $\beta$ is
constant, this results in
\begin{equation*}
\langle v_r^2r\rangle=\frac{GM_\mathrm{tot}}{4(2-\beta)},
\end{equation*}
which differs from the point-mass case in
equation (\ref{eq:kep}) by an exact factor of two.
If the line-of-sight velocity dispersion is used instead, we find
\begin{equation}\label{eq:scvl}
M_\mathrm{tot}
=\frac{12}G\left\langle\frac{2-\beta}{3-2\beta}\,v_\ell^2r\right\rangle
\end{equation}
in place of equation (\ref{eq:vlos2}).

The calculation for the weighting function suitable for the projected
separation as a observable is essentially identical to that found in
Sect.~\ref{sec:prosep},
as we would like to find the weighting function $w(R)$ satisfying
$\langle v_\ell^2w(R)\rangle=\langle(4-2\beta)v_r^2r\rangle$,
which is to be substituted in equation (\ref{eq:scc}). This results in
the identical integral equation (\ref{eq:gie}) with $h=2(2-\beta)r$ or
equivalently $\tilde\mu=1$. However, because of
an additional factor of two in equation (\ref{eq:scc}),
the projected mass estimator for the self-consistent
system with constant $\beta$ is different from equation (\ref{eq:BT})
again exactly by a factor of two, i.e.,
\begin{equation}\label{eq:HTB}
M_\mathrm{tot}=
\frac{64}\pi\frac{2-\beta}{4-3\beta}
\frac{\langle v_\ell^2R\rangle}G,
\end{equation}
which encompasses the result of \citet*{He85}.

\section{Discussion and Conclusions}
\label{sec:dis}

Here, we have developed the theory of mass estimators. We are
motivated by instances in astrophysics in which we wish to estimate
the mass of a dark halo from positions and velocities of tracers such
as stars, globular clusters, and satellite galaxies. The data sets may
then be true distances and radial velocities (as for estimating the
mass of the Milky Way from its satellite galaxies) or may be projected
distances and line-of-sight velocities (as for the local dwarf spheroidal
galaxies). In either case, we wish to estimate the mass of the dark
halo from the kinematics of the tracer population.

For a given halo density profile, there exists the optimal weighting
of these kinematic data. We have shown how to find it for any given
specific density law and different kinds of positional and velocity
data. This means that the mass within any radius can be calculated as
the weighted sum of positions and velocities. We have worked out the
formulae explicitly for a number of important cases, including
scale-free, Hernquist and NFW haloes.

Although we have concentrated on general theoretical developments in
this paper, the performance of the particular scale-free estimators
has been tested and was already reported in an earlier paper
\citep{WEA}.
We have also verified that our estimators (including the NFW ones)
work well against simulation data -- in which a variety of effects
such as halo asphericity, late infall of accreted material and lack of
virial equilibrium are present. Even though these are not taken into
account explicitly in our estimators, none the less they fare well
against simulation data \citep{De10,Ev10}.
Finally, applications of our theory to estimate the masses of the
Milky Way and Andromeda galaxies \citep{WEA} and the dwarf spheroidals
\citep{Ev10} are presented elsewhere.

One notable feature of the mass estimator theories that have been
developed here is that they do not explicitly depend on any
properties of the density of the tracer population $\nu$. In fact,
the mass estimator incorporates the information through the
definition of the average, that is,
\begin{equation*}
\langle u(\bmath r,\bmath v)\rangle\equiv
\frac1{N_\mathrm{tot}}
\int\!\dm^3\!\bmath r\,\dm^3\!\bmath v\,
f(\bmath r,\bmath v)\,u(\bmath r,\bmath v)
=\frac1{N_\mathrm{tot}}\int\!\dm^3\!\bmath r\,\nu(\bmath r)\,\bar u(\bmath r)
\end{equation*}
where
\begin{equation*}\begin{split}
\bar u(\bmath r)&\equiv
\frac1{\nu(\bmath r)}\int\!\dm^3\!\bmath v\,
f(\bmath r,\bmath v)\,u(\bmath r,\bmath v)
\\
\nu(\bmath r)&\equiv\int\!\dm^3\!\bmath v\,f(\bmath r,\bmath v)
\,;\qquad
{N_\mathrm{tot}}\equiv\int\!\dm^3\!\bmath r\,\dm^3\!\bmath v\,
f(\bmath r,\bmath v)
=\int\!\dm^3\!\bmath r\,\nu(\bmath r)
\end{split}\end{equation*}
and $f(\bmath r,\bmath v)$ is the phase space distribution function.
However, given that the sampling of the tracers is statistically
random, this formal average is estimated by the sample mean,
\begin{equation*}
\langle u(\bmath r,\bmath v)\rangle\approx
\frac1N\sum_i^Nu(\bmath r_i,\bmath v_i).
\end{equation*}
That is to say, the effect of the tracer density in our mass estimator
theories is naturally accounted through the spatial frequency of sampled
tracers, and leaves no explicit dependence on $\nu$ in the consequent
formulae. 
In practice, the choice of the sampled tracers may not necessarily be
random. If there is some compelling reason to suspect sampling
bias and/or the specific selection function is known, the sample mean
can be estimated using any additional weighting accounting or
correcting for the selection bias.
However, even if a mild selection bias were to be present,
the mass estimator may reasonably be robust without any explicit
correction. This would be the case if the spatial variation of the
quantity to be averaged for the mass estimator is not strongly
correlated with the variation of the tracer density itself.

The conventional way in which mass estimation is performed is via the
Jeans equations. The procedure is normally as follows: first, an
assumption is made as to the luminosity density of the tracer
population; secondly, the data set of discrete velocities is binned and
smoothed to give the variation of the line-of-sight velocity
dispersion with radius; thirdly, an assumption as to anisotropy is
made (often that the anisotropy parameter $\beta\approx0$) so that
the line-of-sight velocity dispersion can be converted to the radial
velocity dispersion, and fourthly the spherical Jeans equation is
used to relate the underlying potential, and hence the enclosed mass,
to the behaviour of the stellar kinematics. It is worth emphasizing
that the results obtained for the matter distribution are often not
robust, as they depend not just on the luminosity profile and the
second velocity moments, but also their gradients.

In some sense, the techniques in this paper discard the wealth of
information contained in the observed data set by taking spatial
integrals over the whole system. However, there can be a number of
advantages of this approach, especially if the number of datapoints
are limited. First, it guards against overinterpreting the data set
which can often happen with use of the Jeans equations. Second, the
binning of the data, and their subsequent smoothing, are not
needed. This is actually a great help, as Jeans modelling requires
derivatives of functions derived from the binned and smoothed
data. Third, the mass estimators are simple, requiring only weighted
sums of positions and velocities, as opposed to solution of (at best)
an ordinary differential equation.

For these reasons, we expect using mass estimators for discrete data
to be a viable alternative to Jeans modelling. In the limits of large
numbers of datapoints, we expect the mass estimators and Jeans
modelling to yield similarly answers. This is borne out by the
calculations of \citet{Ev10} for dwarf spheroidals where data sets of
thousands of radial velocities are available. When the the number of
tracer datapoints is small, as often happens for estimating halo
masses from satellite galaxies, mass estimators are the technique of
choice. The precious data need not be smoothed, and the
estimates of the enclosed mass are robust. We hope our techniques to
be widely used in these instances, as Jeans modelling is either too
cumbersome or too elaborate.

\section*{acknowledgments}

This work originated from the first author's two separate two-week
visits (February and October 2010) to the IoA (Cambridge), which were
in part supported by the National Natural Science Foundation of China
(NSFC) Research Fund for International Young Scientists, as well as
the IoA's visitor grant. The authors also thank Laura L. Watkins and
Alis J. Deason for Monte Carlo tests concerning the performance of the
mass estimators with the Hernquist and the NFW haloes. JHA is
supported by the Chinese Academy of Sciences (CAS) Fellowships for
Young International Scientists, Grant No.:2009Y2AJ7.

\begin{appendix}

\section{How do we invert for the projection weighting function?}
\label{sec:app}

First, let us rearrange equation (\ref{eq:gie}) into an equivalent
form;
\begin{equation}\label{eq:gie2}
r^3h=\frac{fr^4}{\tilde\mu}
=\int_0^r(r^2-R^2)^{1/2}w(R)\,R\,\dm R
+(1-\beta)\,G(r)
\end{equation}
where
\begin{equation}\label{eq:G}
G(r)\equiv\int_0^r\frac{w(R)\,R^3\dm R}{(r^2-R^2)^{1/2}}.
\end{equation}
However, we find that
\begin{equation}
\frac\dm{\dm r}\left[\frac1r\int_0^r(r^2-R^2)^{1/2}w(R)\,R\,\dm R\right]
=\frac{G(r)}{r^2}.
\end{equation}
Therefore, differentiating equation (\ref{eq:gie2}) after dividing it
by $r$ leads to a differential equation for $G(r)$;
\begin{equation}\label{eq:Gde}
(1-\beta)\frac{\dm G}{\dm r}
+\left(\frac\beta r-\frac{\dm\beta}{\dm r}\right)\,G(r)
=r\frac\dm{\dm r}\biggl(\frac{fr^3}{\tilde\mu}\biggr).
\end{equation}
Given $\tilde\mu(r)$ and $\beta(r)$, it is straightforward to solve
equation (\ref{eq:Gde}) numerically with the boundary condition $G(0)=0$. 
Furthermore, equation (\ref{eq:Gde}) can be brought to an exact form
\begin{equation}\label{eq:ex}
\frac\dm{\dm r}(\mathcal QG)
=\frac{r\,\mathcal Q}{1-\beta}
\frac\dm{\dm r}\biggl(\frac{fr^3}{\tilde\mu}\biggr),
\end{equation}
by means of the integrating factor $\mathcal Q(r)$ satisfying
\begin{equation*}
\frac{\dm\ln\mathcal Q}{\dm r}
=-\frac r{1-\beta}\frac\dm{\dm r}\biggl(\frac\beta r\biggr).
\end{equation*}
Hence, if $\mathcal Q(r)$ can be found, it is even possible to bring
$G(r)$ to a quadrature. Finally, once $G(r)$ is found by some means,
the weighting function $w(R)$ can be obtained though the inverse Abel
transformation of equation (\ref{eq:G}), i.e.,
\begin{equation}\label{eq:ginv}
w(R)=\frac2{\pi R^2}
\int_0^R\frac{\dm r}{(R^2-r^2)^{1/2}}\frac{\dm G}{\dm r}.
\end{equation}

As an example, if $\beta$ is a constant, then $\mathcal Q=r^s$ where
$s=\beta/(1-\beta)$ and thus
\begin{equation}
G(r)
=\frac1{1-\beta}\frac1{r^s}\int_0^r\!\dm\hat r\,\hat r^{s+1}
\left.\frac\dm{\dm r}
\biggl(\frac{fr^3}{\tilde\mu}\biggr)\right|_{r=\hat r}.
\end{equation}
The weighting function $w(R)$ is then found to be
\begin{equation}
w(R)=\frac2{\pi(1-\beta)R^2}
\int_0^R\frac{r\,\dm r}{(R^2-r^2)^{1/2}}
\frac{\dm^2}{\dm r^2}\biggl[\frac1{r^s}\!
\int_0^r\frac{f(\hat r)\hat r^{3+s}\dm\hat r}{\tilde\mu(\hat r)}\biggr].
\end{equation}
For the isotropic case ($\beta=0$), this simplifies to
\begin{equation}
w(R)=\frac2{\pi R^2}\int_0^R\frac{r\,\dm r}{(R^2-r^2)^{1/2}}
\frac\dm{\dm r}\biggl[(3+\hat\alpha)\frac{r^3}{\tilde\mu}\biggr]
\end{equation}
where $\hat\alpha=1-(\dm\ln\tilde\mu/\dm\ln r)$. Similarly for
$\beta=\frac12$ (i.e., $s=1$), the weighting function is found to be
\begin{equation}
w(R)=\frac4{\pi R^4}\int_0^R\frac{r^3\,\dm r}{(R^2-r^2)^{1/2}}
\frac\dm{\dm r}\biggl[(2+\hat\alpha)\frac{r^3}{\tilde\mu}\biggr].
\end{equation}
For a system with purely radial orbits ($\beta=1$), equation
(\ref{eq:ex}) and those derived from it are not valid. However, it is
still possible to solve for $w(R)$ from equation (\ref{eq:gie}) or
(\ref{eq:gie2}), which result in
\begin{equation}
w(R)=\frac2{\pi R^2}\int_0^R\frac{r\,\dm r}{(R^2-r^2)^{1/2}}
\frac{\dm^2}{\dm r^2}\biggl[(1+\hat\alpha)\frac{r^4}{\tilde\mu}\biggr].
\end{equation}
The result for the purely circular orbit cases ($\beta=-\infty$) on
the other hand is obtained by inverting equation (\ref{eq:gcir}) such
that
\begin{equation}\label{eq:pcir}
w(R)=\frac4{\pi R^2}\int_0^R\frac{\dm r}{(R^2-r^2)^{1/2}}
\frac\dm{\dm r}\biggl(\frac{r^4}{\tilde\mu}\biggr).
\end{equation}

\section{The projected mass estimator for the Hernquist halo}
\label{app:hern}

Let us think of
\begin{equation*}\begin{split}
J&=\left\langle\biggl(
\frac{16}\pi\frac{2-\beta}{4-3\beta}\frac R{r_0}+3
+\frac4\pi\frac{1-\beta}{2-\beta}\frac{r_0}R
\biggr)\,v_\ell^2\right\rangle
\\&=\frac{4\pi}{N_\mathrm{tot}}\int_0^\infty\!\dm r\,r^2\nu\sigma_r^2
\left(\frac{16}\pi\frac{2-\beta}{4-3\beta}\frac r{r_0}\mathcal S_2
+3\mathcal S_1
+\frac4\pi\frac{1-\beta}{2-\beta}\frac{r_0}r\mathcal S_0\right)
\end{split}\end{equation*}
where
\begin{equation*}
\mathcal S_n=\int_0^{\pi/2}\!\dm\theta\,\sin^n\!\theta\,(1-\beta\sin^2\!\theta).
\end{equation*}
We find that
\begin{equation*}
\mathcal S_2=\frac\pi{16}(4-3\beta)
\,;\quad
\mathcal S_1=1-\frac23\,\beta
\,;\quad
\mathcal S_0=\frac\pi4(2-\beta).
\end{equation*}
Next,
\begin{multline*}
\frac{16}\pi\frac{2-\beta}{4-3\beta}\frac r{r_0}\mathcal S_2
+3\mathcal S_1+\frac4\pi\frac{1-\beta}{2-\beta}\frac{r_0}r\mathcal S_0
\\=\frac1{2r_0r^{2-2\beta}}
\frac\dm{\dm r}\left[r^{2-2\beta}(r_0+r)^2\right].
\end{multline*}
Therefore
\begin{equation*}\begin{split}
J&=\frac{2\pi}{r_0N_\mathrm{tot}}\int_0^\infty\!dr\,r^{2\beta}\nu\sigma_r^2
\frac\dm{\dm r}\left[r^{2-2\beta}(r_0+r)^2\right]
\\&=
-\frac{2\pi}{r_0N_\mathrm{tot}}\int_0^\infty\!dr\,r^{2-2\beta}(r_0+r)^2
\frac\dm{\dm r}\left(r^{2\beta}\nu\sigma_r^2\right).
\end{split}\end{equation*}
With the spherical Jeans equation for a constant $\beta$ (eq.~\ref{eq:jeans}
with $Q=r^{2\beta}$) and the mass profile for the Hernquist halo
(eq.~\ref{eq:dpm} with $\gamma=p=1$), we find that
\begin{equation*}
J=\frac{2\pi}{r_0N_\mathrm{tot}}
\int_0^\infty\!\dm r\,r^2\nu\,GM_\mathrm{tot}
=\frac{GM_\mathrm{tot}}{2r_0}.
\end{equation*}

Similar calculations can also demonstrate the existence of a rational
projected mass estimator for the particular mass models in equation
(\ref{eq:dpm}) such that
\begin{equation*}
\frac{GM_\mathrm{tot}}{r_0}=\left\langle\biggl(
\frac{32}\pi\frac{2-\beta}{4-3\beta}\frac R{r_0}+3
\biggr)\,v_\ell^2\right\rangle
\end{equation*}
for the Jaffe model
\begin{equation*}
\rho(r)=\frac{M_\mathrm{tot}}{4\pi}\frac{r_0}{r^2(r_0+r)^2}
\,;\qquad
\frac{M(r)}{M_\mathrm{tot}}=\frac r{r_0+r},
\end{equation*}
and
\begin{equation*}
\frac{GM_\mathrm{tot}}{r_0}=\left\langle\biggl(
\frac{32}\pi\frac{2-\beta}{4-3\beta}\frac R{r_0}
+\frac8{\pi}\frac{1-\beta}{2-\beta}\frac{r_0}R
\biggr)\,v_\ell^2\right\rangle
\end{equation*}
for
\begin{equation*}
\rho(r)=\frac{M_\mathrm{tot}}{2\pi}\frac{r_0^2}{r(r_0^2+r^2)^2}
\,;\qquad
\frac{M(r)}{M_\mathrm{tot}}=\frac{r^2}{r_0^2+r^2}.
\end{equation*}
An analytic example for the cored models that require the Dirac delta
in the weighting function is found for
\begin{equation*}
\rho(r)=\frac{3M_\mathrm{tot}}{4\pi}\frac{r_0}{(r_0+r)^4}
\,;\qquad
\frac{M(r)}{M_\mathrm{tot}}=\left(\frac r{r_0+r}\right)^3
\end{equation*}
whose formal form of the projected mass estimator is given by
\begin{equation*}
\frac{GM_\mathrm{tot}}{r_0}=\left\langle\biggl[
\frac{32}\pi\frac{2-\beta}{4-3\beta}\frac R{r_0}+9
+\frac{24}\pi\frac{1-\beta}{2-\beta}\frac{r_0}R
+2(1-2\beta)r_0^2\delta(R^2)
\biggr]\,v_\ell^2\right\rangle.
\end{equation*}

\section{Purely Circular Orbits}
\label{app:cir}

For an extreme scenario, one can imagine that all tracers are in
circular orbits and their orbital phases and orientations are
completely random (hence there is no net angular momentum of tracer
populations in each shell of a fixed radius). This corresponds to the
case that $\beta=-\infty$ everywhere. Since $\sigma_r^2=0$ everywhere,
the $v_r$-based mass estimator is invalid for this case, but the
line-of-sight velocity based ones are still applicable. Although the
final result turns out to be the same as the simple limit to
$\beta\rightarrow-\infty$, we can derive them via physically consistent
routes.

Let us start by noting that the line-of-sight velocity dispersion of
the population in purely circular orbits with random orientations is
given by
\begin{equation*}
\sigma_\ell^2=\frac{v_\mathrm c^2}2\sin^2\!\theta
=\frac{GM(r)}{2r}\sin^2\!\theta
\end{equation*}
where $v_\mathrm c$ is the circular speed of the spherical halo at $r$.
First, we consider the case that the radial distances $r$ of the
individual tracers to the halo centre are known. Then, the average of
$v_\ell^2$ weighted by a function $h(r)$ is found to be
\begin{equation}\label{eq:vtan}\begin{split}
\left\langle v_\ell^2h(r)\right\rangle&=\frac{4\pi}{N_\mathrm{tot}}
\int_{r_\mathrm{in}}^{r_\mathrm{out}}\!\dm r
\int_0^{\pi/2}\!\dm\theta\,r^2\sin\theta\,\nu\sigma_\ell^2h(r)
\\&=\frac{4\pi}{N_\mathrm{tot}}
\int_{r_\mathrm{in}}^{r_\mathrm{out}}\!\dm r\,r^2\nu\frac{v_\mathrm c^2h}2
\int_0^{\pi/2}\!\dm\theta\,\sin^3\!\theta
\\&=\frac{4\pi}{N_\mathrm{tot}}
\int_{r_\mathrm{in}}^{r_\mathrm{out}}\!\dm r\,r^2\nu\frac{GMh}{3r}.
\end{split}\end{equation}
Hence, if one chooses $h=3r/\tilde\mu$ where $\tilde\mu(r)$ is the
assumed mass profile (eq.~\ref{eq:mprof}), the total mass
$M_\mathrm{out}$ can be isolated by
\begin{equation}\label{eq:vltme}
M_\mathrm{out}=\frac3G\left\langle\frac{v_\ell^2r}{\tilde\mu}\right\rangle,
\end{equation}
which is consistent with equation (\ref{eq:vlos2}) in the limit of
$\beta\rightarrow-\infty$. The result does not involve the boundary
term because $\sigma_r^2=0$ everywhere for the assumed system.

For the case that only the projected distances $R$ to the halo centre
are available, we consider the similar weighted average of $v_\ell^2$
as Sect.~\ref{sec:prosep}, that is,
\begin{equation*}
\left\langle v_\ell^2w(R)\right\rangle
=\frac{4\pi}{N_\mathrm{tot}}\int\!\dm r\,r^2\nu\frac{GM}{2r}
\int_0^{\pi/2}\!\dm\theta\,\sin^3\!\theta\,w(R).
\end{equation*}
Hence, if $w(R)$ is chosen to satisfy the integral equation
\begin{equation}
\label{eq:gcir}
\frac{2r^4}{\tilde\mu}=\int_0^r\frac{w(R)\,R^3\dm R}{(r^2-R^2)^{1/2}}
\equiv G(r),
\smallskip\end{equation}
the total mass is related to the weighted average via
$GM_\mathrm{out}=\langle v_\ell^2w(R)\rangle$.
For the scale-free case (eq.~\ref{eq:musf}), we have
$w(R)\propto R^\alpha$ and therefore 
\begin{equation*}
M_\mathrm{out}
=\frac{4\Gamma\bigl(\tfrac{\alpha+5}2\bigr)}
{\pi^{1/2}\Gamma\bigl(\tfrac\alpha2+2\bigr)}
\frac{\langle v_\ell^2R^\alpha\rangle}
{Gr_\mathrm{out}^{\alpha-1}},
\end{equation*}
which is valid for $\alpha>-4$. For a central point mass ($\alpha=1$),
this becomes $GM_\mathrm{tot}=\frac{32}{3\pi}\langle v_\ell^2R\rangle$.
That is to say, the mass estimate under the assumption that the
tracers in purely circular orbits is smaller by a factor of three and
1.5 respective compared to the case that they are in the radial orbits
or the isotropic case (see eq.~\ref{eq:BT}).

The calculations for the self-consistent case are similar. First, we
use $\nu/N_\mathrm{tot}=\rho/M_\mathrm{tot}$ (assuming
$r_\mathrm{in}=0$ and $r_\mathrm{out}=\infty$) and
$\dm M/\dm r=4\pi r^2\rho$ to further reduce equation (\ref{eq:vtan}) to
\begin{equation*}
\left\langle v_\ell^2h(r)\right\rangle=\frac1{M_\mathrm{tot}}
\int_0^\infty\!\dm r\,\frac{Gh}{6r}\frac{\dm M^2}{\dm r}.
\end{equation*}
With $h=6r$, we find that
$M_\mathrm{tot}=6G^{-1}\langle v_\ell^2r\rangle$, which differs from
equation (\ref{eq:vltme}) with $\tilde\mu=1$ by a factor of two and is
also the limit of equation (\ref{eq:scvl}) as $\beta\rightarrow-\infty$.
With $R$ as an observable instead of $r$, we have $G(r)=4r^4$ in place
of equation (\ref{eq:gcir}), which ultimately leads to the mass estimator,
$GM_\mathrm{tot}=\frac{64}{3\pi}\langle v_\ell^2R\rangle$, which is
the same as equation (\ref{eq:HTB}) for $\beta\rightarrow-\infty$.

\end{appendix}

\label{lastpage}
\end{document}